# High entropy effect on thermoelectric properties of nonequilibrium cubic phase of $AgBiSe_{2-2x}S_xTe_x$ with $x$ = 0–0.6

Asato Seshita,[a] Aichi Yamashita,[a,*] Takayoshi Katase,[b] Yoshikazu Mizuguchi[a]

[a] *Department of Physics, Tokyo Metropolitan University, 1-1 Minami-Osawa, Hachioji-shi, Tokyo 192-0397, Japan*

[b] *MDX Research Center for Element Strategy, International Research Frontiers Initiative, Tokyo Institute of Technology, 4259 Nagatsuta, Midori, Yokohama 226-8501, Japan.*

**Abstract**

Silver bismuth diselenide ($AgBiSe_2$) has much attention as an efficient thermoelectric material due to its low thermal conductivity. However, $AgBiSe_2$ exhibits multiple crystal structural transitions with temperature, and high thermoelectric performance was realized only in high-temperature cubic phase. We previously reported the stabilization of cubic phase in $AgBiSe_{2-2x}S_xTe_x$ with $x$ = 0.6–0.8 at room temperature by high-entropy-alloy (HEA) approach. The cubic HE-type $AgBiSe_{0.8}S_{0.6}Te_{0.6}$ achieved a high *ZT* value of 0.8 at 748 K. In this paper, we succeeded in stabilizing the cubic phase in $AgBiSe_{2-2x}S_xTe_x$ with $x$ = 0–0.6 by ice-quenching method, and investigated the HE effect on the thermoelectric properties. Cubic $AgBiSe_{2-2x}S_xTe_x$ exhibited n-type conductivity from 300 K to 10 K. We found that electronic conductivity was largely increased around room temperature with increasing the amount of S and Te, although carrier concentration showed almost the same values. The S and Te substitutions induced the variation of band structure, resulting in the carrier mobility enhancement. Furthermore, thermal conducutivity showed reduction tendency with increasing the amount of S and Te due to enhancement of phonon scattering. Simultaneous electronic conductivity increase and thermal conductivity reduction resulted in the systematic improvement of the *ZT* values for HE-type cubic $AgBiSe_{2-2x}S_xTe_x$.

*Corresponding author: Aichi Yamashita

E-mail: aichi@tmu.ac.jp

## Introduction

The thermoelectric generators can play an important role in a sustainable energy solution since they can directly convert heat into electricity. Thermoelectric energy conversion efficiency is defined as figure of merit $ZT$; $ZT = S^2T/\rho\kappa_{tot}$, where $S$ is Seebeck coefficient, $T$ is absolute temperature, $\rho$ is electrical resistivity, and $\kappa_{tot}$ is total thermal conductivity, respectively. The $\kappa_{tot}$ is the summation of lattice ($\kappa_{lat}$) and electronic ($\kappa_{ele}$) contributions. The term of $S^2/\rho$, called as power factor (PF), can be maximized by carier concentration and band tuning. Therefore, high-performance thermoelectric materials require high PF and low $\kappa_{tot}$, as shown in the formula of $ZT$.[1]

Recently, silver bismuth diselenide ($AgBiSe_2$) has attracted much interest as a high-performance thermoelectric material due to its intrinsically low $\kappa_{tot}$ < 1 $Wm^{-1}K^{-1}$ at room temperature.[2–13] Both n-type and p-type thermoelectric properties of $AgBiSe_2$ were reported. The difference in polarity of $AgBiSe_2$ is due to the defect of Se.[8] Nb-doped[3] and Cl-doped[4] $AgBiSe_2$ bulks achieved high $ZT$ of 1 and 0.9, respectively. The remarkable characteristic of $AgBiSe_2$ is structural phase transitions with temperature, as shown in Fig. 1. It crystallizes in a hexagonal $P\bar{3}m1$ (#164) space group at room temperature, a rhombohedral $R\bar{3}m$ (#166) space group around 490 K, and a cubic $Fm\bar{3}m$ (#225) space group above 590 K. However, this complicated crystal structural transition would be a drawback for module application due to the difference in crystal structure between low-temperature and high-temperature sides. This transition issue was solved by our previous work using the concept of high-entropy-alloys (HEAs).[13] HEAs are typically defined as — (a) alloys containing five or more elements with a concentration between 5 and 35%, (b) 1.5$R$ or more mixing entropy $\Delta S_{mix}$, calculated by $\Delta S_{mix} = -R\Sigma_i c_i \ln c_i$, where $c_i$ and $R$ are the compositional ratio and the gas constant, respectively.[14,15] HEAs introduce a new path of developing advanced materials with unique properties such as severe lattice-distortion effect, fatigue strength, and sluggish diffusion effect. The HEAs have been developed for "metal alloys"; on the other hand, many high-entropy-type (HE-type) compounds, for which the HEAs concept is extended, have been reported recently. For instance, new HE-type layered compounds and NaCl-type metal chalcogenides as superconductors and thermoelectric materials have been reported.[13,16–24] Focused on thermoelectric materials, HE-type compounds are expected to — (a) the stabilization of solid solution phases by decreasing Gibbs free energy with an increase in the entropy, (b) the reduction of the $\kappa_{lat}$ by enhancing lattice disorder.[25] Recently we have reported new HE-type $AgBiSe_{2-2x}S_xTe_x$ with ultra-low $\kappa_{lat}$ and maximum $ZT$ value of 0.9 without optimizing carrier concentration.[26] The cubic phase was successfully stabilized at room temperature with above $x = 0.6$. According to the analysis of $\kappa_{lat}$ in the hexagonal phase, the point defects scattering (PDS) model estimated with the contribution of mass and strain contrast indicates the lattice disorder and/or strain by S and Te substitution fostered the phonon scattering and contributed to the decrease of $\kappa_{lat}$. However, the S and Te substitution effects on the thermoelectric properties including the mechanism of the decrease of $\kappa_{lat}$ in the cubic phase could not be discussed in whole $x$ amount due to the appearance of the impurity phase at high temperature and incompletely stabilized cubic phase between $x = 0$–$0.5$ at room temperature. In the present work, we successfully stabilized the cubic phase of HE-type $AgBiSe_{2-2x}S_xTe_x$ with $x = 0$–$0.6$ by conducting the ice-quenching at the enough high temperature for single phase of cubic structure,[26] and investigated the S and Te substitution effects on the thermoelectric properties below room temperature.

**Experimental**

Polycrystalline samples of $AgBiSe_{2-2x}S_xTe_x$ with $x$ = 0, 0.2, 0.4, 0.6 were prepared by melting method using elemental grains of Ag (99.99%, Kojundo Kagaku), Bi (99.999%, Kojundo Kagaku), Se (99.999%, Kojundo Kagaku), S (99.9999%, Kojundo Kagaku), and Te (99.999%, Kojundo Kagaku). The stoichiometric ratio of grains was annealed in an evacuated quartz tube at 1,000°C for 15 h, and then the obtained powders were pelletized and heated at 500°C for 20 h. Hot-press treatment was performed in order to densify the obtained compounds using graphite dies at 500°C for 30 min under a uniaxial pressure of 60 MPa. Samples in the form of rectangular solids (typically a 1.5-mm-width, 2.5-mm-depth, 80-mm-hight) were cut from the cylinders using a diamond saw for electrical and thermal transport property measurements. The samples were placed in the evacuated quartz tube and heated at 450°C, which all samples become cubic phase, for 5 h. Then ice-quench treatment was performed to obtain the single phase of cubic structure. Relative density of all samples was beyond 90%, and there was no large difference in the relative density between the as-annealed and ice-quenched samples.

The actual composition was analyzed by energy-dispersive X-ray spectroscopy (EDX) on a TM-3030 (Hitachi Hightech) equipped with an EDX-SwiftED analyzer (Oxford). The phase purity and crystal structure were examined by powder X-ray diffraction (XRD), using Miniflex-600 (RIGAKU) with $CuK_\alpha$ radiation in Bragg-Brentano $\theta$-$2\theta$ method with a high-resolution semiconductor detector/tex-Ultra at room temperature. The crystal structure parameters were refined using the Rietveld method using a RIETAN-FP software,[27] and crystal structures were visualized using VESTA software.[28]

Temperature dependence of electrical resistivity ($\rho$), Seebeck coefficient ($S$), and total thermal conductivity ($\kappa_{tot}$) were measured on the samples of $AgBiSe_{2-2x}S_xTe_x$ with $x$ = 0, 0.2, 0.4, 0.6 from room temperature down to 10 K. The $\rho$ was measured using the four-probe method with the electrical resistivity measurement option of Magnetic Property Measurement System (MPMS3, Quantum Design). A Thermal Transport Option (TTO) of Physical Property Measurement System (PPMS,

Quantum Design) was used for the $\kappa_{tot}$ and the $S$ measurements with the two-probe method. The Hall resistivity was measured with PPMS under magnetic field up to 5 T to examine the Hall coefficient $R_H$.

Sound velocity measurement was performed using 1077DATA (KARL DEUTSCH) for longitudinal $v_{long}$ and transverse sound velocity $v_{trans}$. The Grüneisen parameter $\gamma$, Poisson ratio $\nu_P$, and mean sound velocity $v_{mean}$ were calculated by following equations.[29]

$$\gamma = \frac{3}{2}\left(\frac{1+\nu_P}{2-3\nu_P}\right)$$

$$\nu_P = \left(1-2\left(\frac{v_{trans}}{v_{long}}\right)^2\right) \Big/ \left(2-2\left(\frac{v_{trans}}{v_{long}}\right)^2\right)$$

The Debye temperature $\Theta_D$ can be written using $v_{mean}$, Dirac's constant $\hbar$, Boltzmann constant $k_B$, and volume per atom $\Omega$.[30,31]

$$\Theta_D = \frac{\hbar}{k_B}\left(\frac{6\pi^2}{\Omega}\right)^{1/3} \cdot v_{mean}$$

$${v_{mean}}^3 = \frac{3}{{v_{long}}^{-3} + 2{v_{trans}}^{-3}}$$

Point defect scattering model for alloy phase can assess the lattice thermal conductivity $\kappa_{lat}$ using following equations.[5,6,32–35]

$$\frac{\kappa_{lat,alloy}}{\kappa_{lat,pure}} = \frac{\arctan(u)}{u}$$

The $\kappa_{lat,alloy}$ is the lattice thermal conductivity of a disordered alloy, and the $\kappa_{lat,pure}$ is the lattice thermal conductivity of a pure crystal without disorder. The disorder scaling factor $u$ can be written using the $\Theta_D$, the $\Omega$, the $v_{mean}$, and the scattering parameter $\Gamma$.

$$u^2 = \frac{\pi\Theta_D\Omega}{2\hbar v_{mean}}\kappa_{lat,pure}\Gamma$$

The $\Gamma$ is the summation of mass contribution $\Gamma_M$ and strain contribution $\Gamma_S$.

$$\Gamma_M = \frac{\sum_{i=1}^{n} c_i \left(\frac{\overline{M}_i}{\overline{\overline{M}}}\right)^2 \sum_k f_i^k \left(1-\frac{M_i^k}{\overline{M}_i}\right)^2}{\left(\sum_{i=1}^{n} c_i\right)}$$

$$\Gamma_{\mathrm{S}} = \frac{\sum_{i=1}^{n} c_i \varepsilon_i \left(\frac{\bar{M}_i}{\bar{\bar{M}}}\right)^2 \sum_k f_i^k \left(1 - \frac{r_i^k}{\bar{r}_i}\right)^2}{\left(\sum_{i=1}^{n} c_i\right)}$$

$\bar{\bar{M}}$ is the average mass of compound, $c_i$ is relative degeneracies of $i$th sublatt $\bar{M}_i$ and $\bar{r}_i$ are average mass and average atomic radius of $i$th sublattice, $f_i^k$ is fractional occupation of $k$th atom of the $i$th sublattice, $M_i^k$ and $r_i^k$ are mass and atomic radius of $k$th atom of the $i$th, respectively. $\varepsilon_i$ is the elastic parameter and regarded as a phenomenological adjustable parameter.[35]

## Results and discussion

### Structural characterization

Figure 2 shows the room-temperature powder XRD patterns for $AgBiSe_{2-2x}S_xTe_x$ with $x$ = 0.0–0.6 before and after quenching. All diffraction peaks for quenched samples were indexed by the rock-salt type structure (space group: $Fm\overline{3}m$, #225) which is well known as the typical high-temperature phase of $AgBiSe_2$,[2–6,8–13] indicating the successful stabilization of cubic phase without impurities at room temperature. The diffraction peaks for $x$ = 0.0–0.4 before quenching showed the hexagonal structure. While the diffraction peaks for $x$ = 0.6 before quenching was attributed to the rock-salt type structure, the peaks become sharper after quenching, indicating the crystallinity is much improved in the quenched sample. To assess the sample homogeneity, SEM imaging and EDX mapping were performed. No secondary phase was found in the SEM images (Fig. S1), and also EDX mapping (Fig. S2) revealed that there was no compositional inhomogeneity in all samples. An actual chemical composition, mixing entropy ($\Delta S_{\mathrm{mix}}$), lattice constant, and volume of the unit cell are summarized in Table 1. The actual chemical composition was estimated using the EDX, and the estimated values were almost same as the nominal compositions, even though tiny deficiencies of chalcogenides were found. The $\Delta S_{\mathrm{mix}}$ was estimated by the following equations.[16] The value of $\Delta S_{\mathrm{mix}}$ at $i$th crystallographically independent (CI) site is given by

$$\Delta S_{\mathrm{mix}}^{i} = -R\sum_{j=1}^{N} x_j \ln x_j$$

where $R$, $N$, and $x_j$ are the gas constant, the number of the component at the $i$th CI site, and the atomic fraction of the component, respectively. This formula is utilized to calculate the total mixing entropy of the unit cell.

$$\Delta S_{\mathrm{mix}}(\mathrm{total}) = \sum_{i=1}^{n} \Delta S_{\mathrm{mix}}^{i}$$

where $n$ is the number of the CI sites in the unit cell. According to the value $\Delta S_{\mathrm{mix}}$ in Table 1, the samples of $x$ = 0.4 and 0.6 can be considered as HE-type compounds. Rietveld refinement was performed to estimate the lattice constant and the volume of unit cell. The results of Rietveld refinement are shown in Fig. S3. The lattice constant systematically increased with $x$, and the behavior corresponds to the increase in average ionic radius of chalcogenide sites with equimolar ratio of S (0.184 nm, coordination number of 6) and Te (0.221 nm, coordination number of 6).

**Electrical transport properties**

The $\rho$ of quenched $AgBiSe_{2-2x}S_xTe_x$ samples is shown in Fig. 3. The magnitude of $\rho$ at room temperature ranged from 52 to 442 mΩcm. Systematic decrease in $\rho$ was observed with increasing $x$. The temperature dependence of $\rho$ showed different behavior from both the typical semiconductors and the previously reported hexagonal structure of $AgBiSe_2$.[3] It can be divided into following three regimes (i– iii) similar to the $(Bi, Sb)_2(Te, Se)_3$:[36,37]

(i) *Activated regime.* In the temperature range from around 100 K to 300 K, the $\rho$ can be fitted with the Arrhenius law,

$$\rho \sim \exp(\Delta/k_B T)$$

where Δ is the activation energy, $k_B$ is the Boltzmann constant. Fig. 3(b) shows the Arrhenius plot of the data with linear fittings (solid line) to estimate Δ. In Table 1, Δ has a negative correlation with $x$ (Fig. S4), indicating that the energy gap between the conduction band and donor level probably becomes narrower because quenched $AgBiSe_{2-2x}S_xTe_x$ samples show n-type conductivity by Hall measurement.

(ii) *Variable-range hopping (VRH) regime.* In the temperature range from ~80 K to ~100 K, the $\rho$ can be described by Mott's 3D-VRH behavior,[38]

$$\rho \sim \exp\left[(T_0/T)^{1/4}\right]$$

where $T_0$ is a constant that depends on the density of state at the Fermi level $E_F$. The linear fittings (dashed line) of VRH behavior are shown in Fig. 3(c). This hopping conduction occurs due to localized states in a narrow band near the $E_F$. The temperature range in which VRH behavior appear overlapped with the activated temperature range. The distinction between the two transport mechanisms depend on the level of disorder in the sample as discussed in the previous paper.[37]

(iii) *Saturation regime.* Below the VRH temperature range, the $\rho$ tended to saturate, rather than to diverge as expected for intrinsic semiconductors. This saturation behavior implies that some extended states and two or more transport channels exist at $E_F$ in the zero-temperature limit, moreover chemical potential is pinned to these states.[37] Antisites between Ag, Bi cation and S, Se, Te anion are assumed as the factor of the existing these states. In $Bi_2Se_3$, the antisites create some shallow donor or acceptor levels;[39–41] hence, antisite should be the reason for this saturation behavior. To reveal the origin of this phenomenon, further information from the experiments, which can observe electronic defect states, such as scanning tunnel microscopy (STM) or current imaging tunneling spectroscopy (CITS).[39] Figure 4(a) shows the temperature dependence of Hall carrier concentration $n_H$. The $n_H$ showed almost the same values, indicating that the $\rho$ behavior around room temperature can be described as Hall mobility contribution. In quenched $AgBiSe_2$, $n_H$ gradually decreases with decreasing temperature and saturates below around 75 K, indicating Hall carrier concentration is almost constant in the saturation regime. The temperature dependence of $n_H$ for other samples exhibited the same behavior, although it was not able to be measured below 50 K because of difficulty in the measurement of reliable Hall voltage. Temperature dependence of Hall mobility $\mu_H$ is shown in Fig. 4(b). The $\mu_H$ of almost all samples showed a reduction tendency with decreasing temperature. In VRH conduction, its constituent states are far away from each other, and their spatial distribution can be considered uncorrelated.[38] Hence, this $\mu_H$ behavior should be the effect of hopping conduction or some transport channels. This kind of behavior was observed in Ni-doped $CoSb_3$ which has VRH conduction in low temperature.[42]

Fig. 5(a) shows the temperature dependence of *S*. Around room temperature, the absolute value of *S* systematically decreases with increasing the amount of *x*. The absolute values of *S* slightly decreased when temperature is reduced to ~200 K, and slightly increased until the range of 50−100 K.

Finally, they dramatically decreased and became close to zero. The temperature region of $S$ behavior corresponds to that of VRH or $\rho$ saturation, and it implies the localized states or pinned extended states. This $S$ behavior is similar to Ni-doped $CoSb_3$ behavior,[42] where the $S$ vanishing is explained by the effect of the impurity band from Ni. The $S$ behavior around room temperature corresponds to the behavior of the $\rho$. However, $n_H$ contribution is less discussed above. The Pisarenko plot at 275 K is shown in Fig. 5(b). They did not follow the SPB model trend, indicating that the reason for decreasing the absolute value of $S$ should be band flattening or multi-band effect.[43,44] This result indicates increasing $\mu_H$ with increasing $x$ is due to the modification of band structure by S and Te substitutions.

**Thermal transport properties**

The $\kappa_{tot}$, $\kappa_{lat}$, and $\kappa_{ele}$ are shown in Fig 6(a), (b), and Fig. S5, respectively. The $\kappa_{ele}$ is directly estimated using Wiedemann-Franz law: $\kappa_{ele} = LT/\rho$, where $L$ is Lorenz number which is expressed by $L = 1.5 + \exp(-|S|/116)$ (where $L$ is in $10^{-8}$ and $S$ in μV/K).[45] The $\kappa_{lat}$ is obtained by subtracting $\kappa_{ele}$ from $\kappa_{tot}$. The $\kappa_{tot}$ for $x = 0$ shows weak temperature dependence up to ~100 K, and largely decreased below ~100 K. Finally, $\kappa_{tot}$ suddenly increased around 20 K. Around the room temperature and 20 K regions, measurement was not stable and they have large error bar as shown in Fig. S6. The $\rho$ and $S$ showed unexpected behavior possibly due to some extended states near $E_F$; therefore, $\kappa_{tot}$ should be also affected by the forementioned matter in the case of ~20 K region. Other samples show similar behavior and $\kappa_{lat}$ almost the same, indicating that the contribution of the $\kappa_{tot}$ is dominant by $\kappa_{lat}$. The $\kappa_{lat}$ is expressed as $1/3Cvl$, where $C$ is lattice specific heat, $v$ is sound velocity, and $l$ is the mean free path of phonons. According to the equation, increase in the $C$ raises the $\kappa_{lat}$ until around one-third of Debye temperature, and $\kappa_{lat}$ almost saturates above that temperature. In general, $\kappa_{lat}$ has a peak around that temperature due to the decreasing $l$ which is caused by phonon-phonon Umklapp scattering.[46] However, the quenched $AgBiSe_{2-2x}S_xTe_x$ samples did not show the peak possibly due to the shortening of $l$ against the temperature by severe phonon scattering systems.[47] The $\kappa_{lat}$ around room temperature almost systematically decreases with increasing $x$. Here, point defect scattering (PDS) model fitting was performed to assess the magnitude of the substitution effect for $\kappa_{lat}$, and the result at 275 K is shown in Fig. 6(c). This model can evaluate the $\kappa_{lat}$ of the contribution of mass and strain contrast. The $\kappa_{lat}$ and mean sound velocity $v_{mean}$ of the pristine sample is required for this model. The $\kappa_{lat}$ of 0.90 W/mK and the $v_{mean}$ of 1636.2 $ms^{-1}$, obtained from a longitudinal sound velocity of 2888.9 $ms^{-1}$ and transversal sound velocity of 1459.4 $ms^{-1}$, were utilized to perform the PDS model. The $\varepsilon$ of 5 gives good fitting for the trend of quenched $AgBiSe_{2-2x}S_xTe_x$. No deviation between the mass contribution and mass and strain contribution in Fig. 6(c) suggests that the mass contribution dominates in this cubic system. The $\varepsilon$ of the hexagonal $AgBiSe_{2-2x}S_xTe_x$ is 32.5, and it shows that the contribution fraction of mass and strain is largely different. This difference should be due to the difference in anharmonicity. The Grüneisen parameter of the hexagonal $AgBiSe_2$ was estimated as 1.68[13], while that of the cubic $AgBiSe_2$ was 1.97, indicating the high anharmonicity nature of the cubic structure of $AgBiSe_2$ system. Moreover, the atomic bond length of the cubic phase is more fluctuated than that of the hexagonal phase for $AgBiSe_2$.[7] Hence, the pristine cubic sample already has a large strain indicating that the contribution of substitution of S and Te for strain should be small.

The power factor PF of quenched $AgBiSe_{2-2x}S_xTe_x$ is shown in Fig. 7(a). The PF for $x = 0$ was 0.54 $\mu Wcm^{-1}K^{-2}$ at room temperature and decreased with the temperature. The maximum PF among the obtained samples were 1.27 $\mu Wcm^{-1}K^{-2}$ for $x = 0.6$ at room temperature. Although PF for $x = 0.4$ and 0.6 showed almost the same values of PF, it tended to increase with increase in the $x$ amount.

Fig. 7(b) shows the *ZT* of quenched $AgBiSe_{2-2x}S_xTe_x$. The *ZT* values systematically increased with $x$ amount, indicating that the simultaneous substitution of Se site by S and Te improved the *ZT* values due to the improvement of both electrical and thermal properties. The maximum *ZT* reached 0.05 with $x = 0.6$ at room temperature.

## Conclusion

In summary, we successfully stabilize the high-temperature cubic phase in $AgBiSe_{2-2x}S_xTe_x$ with $x$ = 0–0.6 at room temperature by ice-quench technique and investigated the HE effects on the thermoelectric properties of the cubic phase. Around room temperature, systematic change of electrical transport properties was observed and it was discussed using SPB model. In consequence, it was explained by the variation of band structure. In lower temperature regions, the $\rho$ showed 3D-VRH and saturated behavior possibly due to some localized states or extended states near $E_F$. Moreover, the vanishment of $S$ around 50 K also could be caused by the same mechanism. The existence of some transport channels is consistent with the result of the Pisarenko plot, suggesting that the simultaneous substitution of S and Te for Se site should affect band structure. The temperature dependence of $\kappa_{lat}$ without a peak around 50 K indicates the shortening of $l$ for $AgBiSe_{2-2x}S_xTe_x$, possibly due to the severe phonon scattering. Finally, the $ZT$ value of 0.05 was achieved for the $x$ = 0.6 sample at room temperature. Our work offered an approach to investigate the thermoelectric properties of cubic structure of $AgBiSe_{2-2x}S_xTe_x$ and we succeeded in enhancing the thermoelectric properties around room temperature.

**Conflicts of interest**

There are no conflicts to declare.

**Acknowledgements**

The authors thank Hiroto Arima for his supports in experiments. This work was partially performed with Collaborative Research Project of Laboratory for Materials and Structures, Institute of Innovative Research, Tokyo Institute of Technology. A. Y. was partly supported by a Grant-in-Aid for Scientific Research (KAKENHI) (No. 22K14480), and the Asahi Glass Foundation. Y. M. was partly supported by a Grant-in-Aid for Scientific Research (KAKENHI) (No. 21H00151), JST-ERATO (No. JPMJER2201), and the Tokyo Metropolitan Government Advanced Research (No. H31-1).

**Figure**

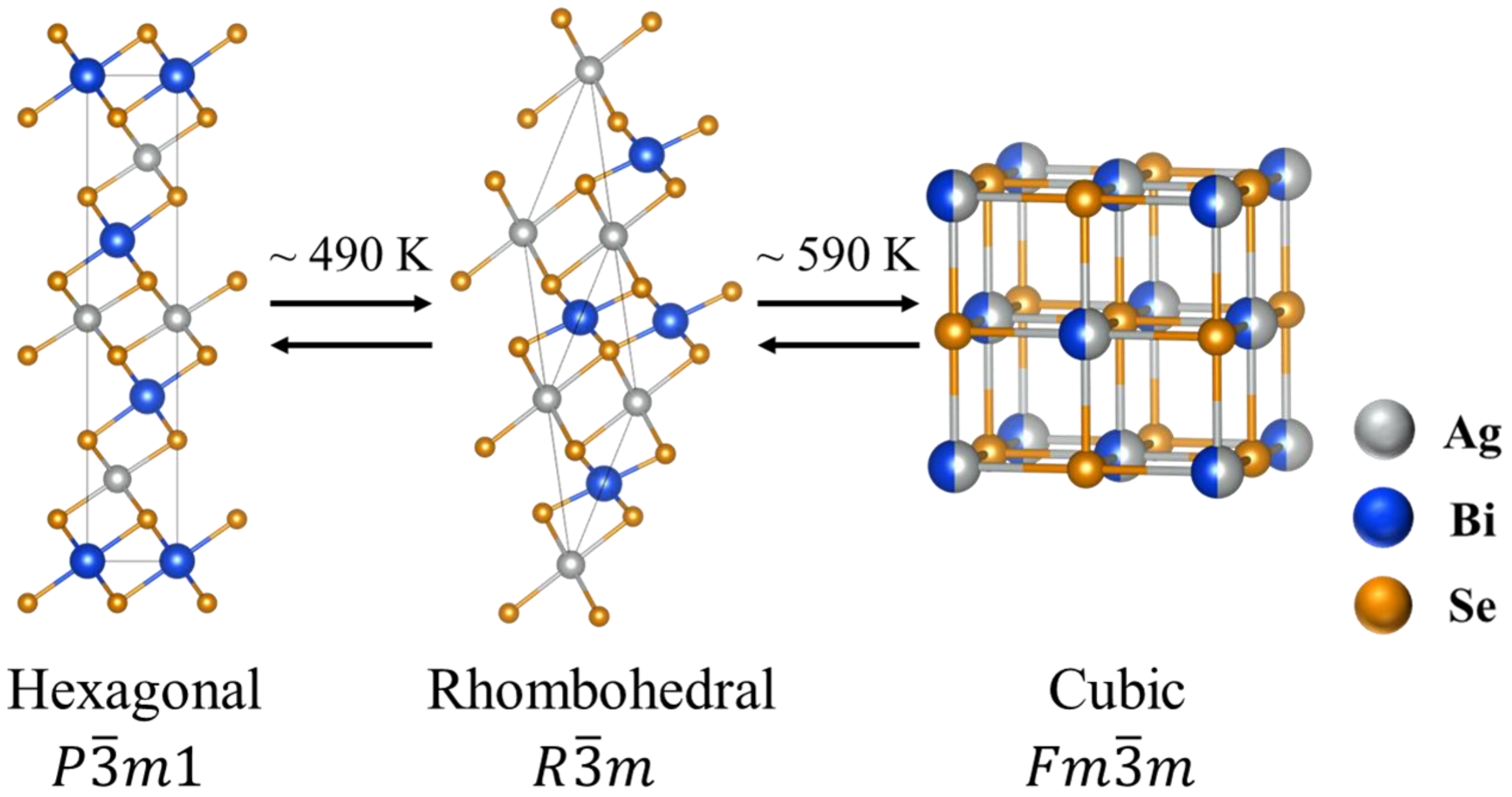

**Fig. 1** Temperature dependence of crystal structure change in $AgBiSe_2$.

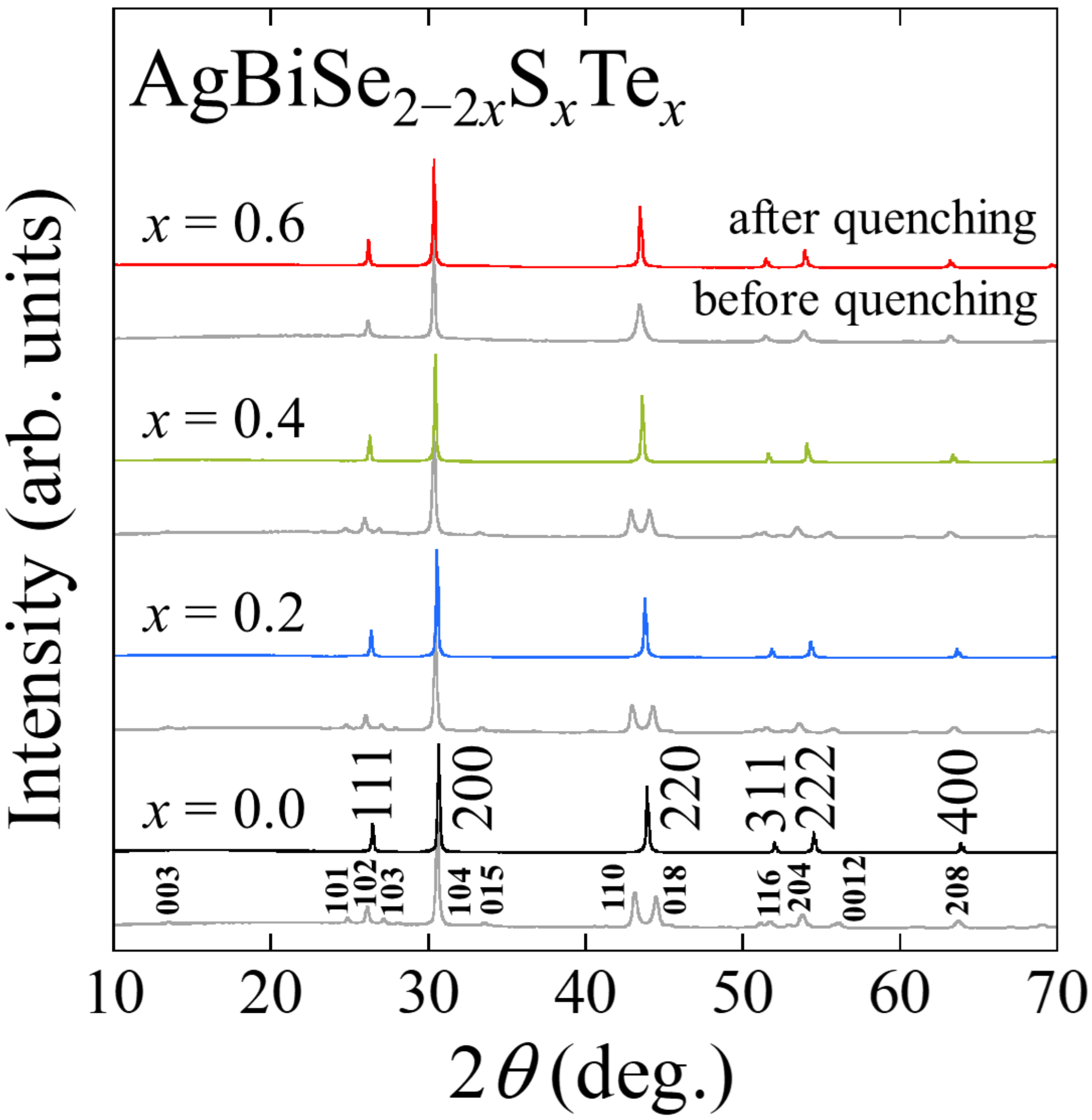

**Fig. 2** X-ray diffraction (XRD) patterns of quenched $AgBiSe_{2-2x}S_xTe_x$ samples at room temperature. XRD patterns of before quenching samples were also plotted below those of each quenched samples as a comparison.

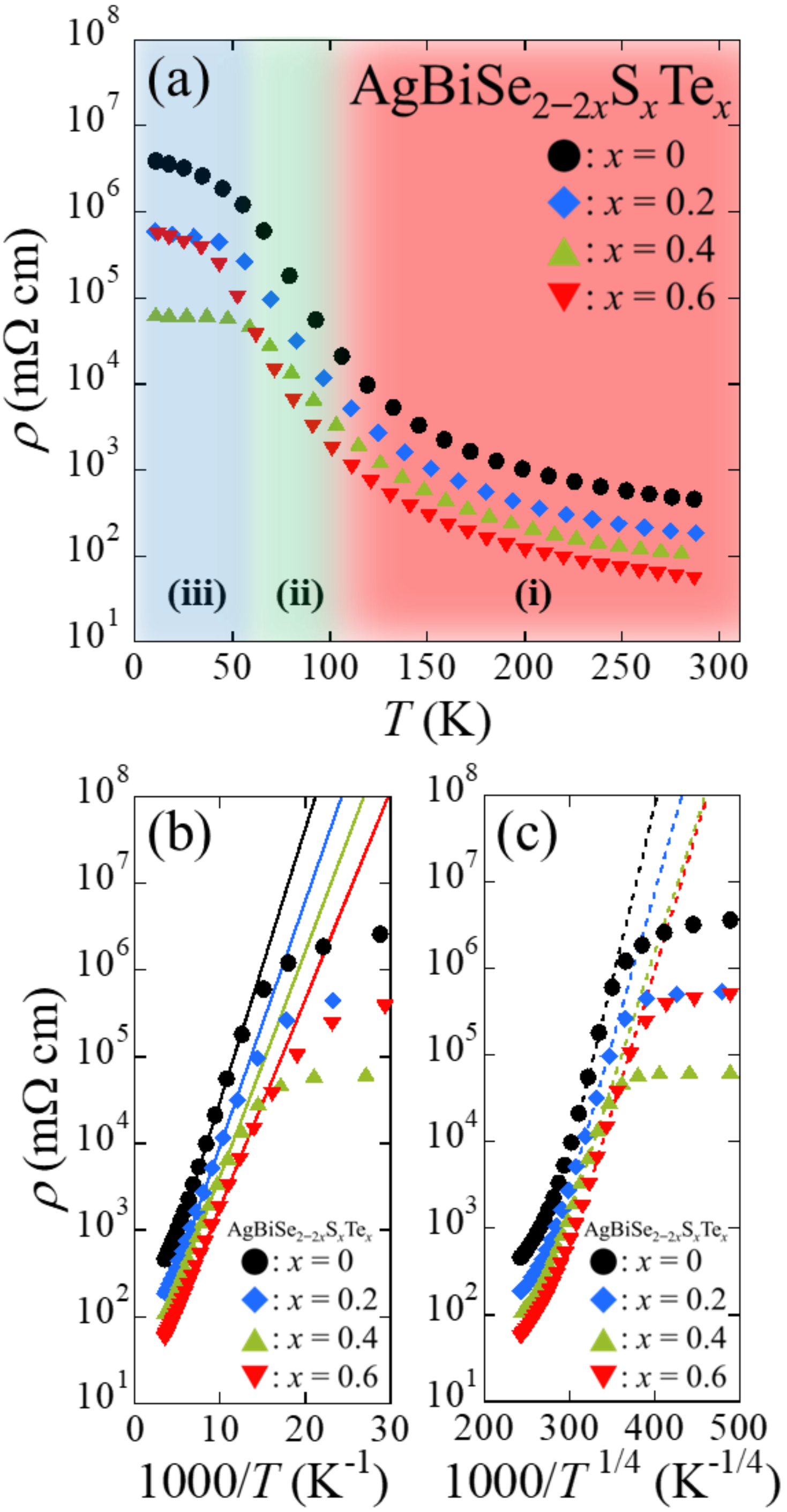

**Fig. 3** Temperature ($T$) dependence of electrical resistivity ($\rho$) for quenched $AgBiSe_{2-2x}S_xTe_x$ samples. Red, green and blue regions are activated regime, variable-range hopping (VRH) regime and saturation regime, respectively. (b) and (c) are the Arrhenius plot [$\log_{10}\rho$ vs $1000/T$] and the 3D- VRH plot [$\log_{10}\rho$ vs $1000/T^{1/4}$] of electrical resistivity data, respectively. Solid line and Dashed line represent linear fittings of the activation and VRH behaviors, respectively.

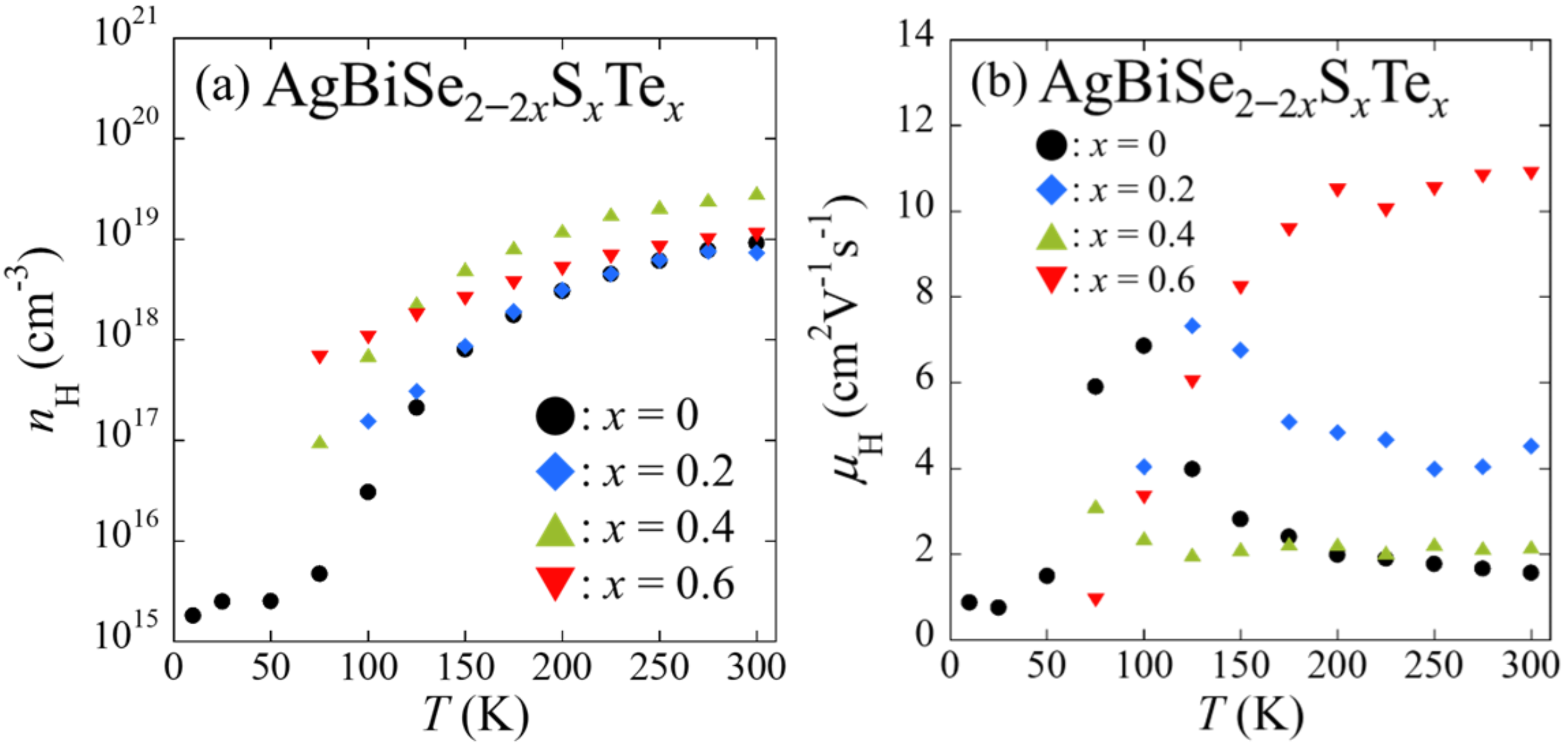

**Fig. 4** Temperature ($T$) dependence of (a) Hall carrier concentration ($n_H$), and (b) Hall mobility ($\mu_H$) for quenched $AgBiSe_{2-2x}S_xTe_x$ samples.

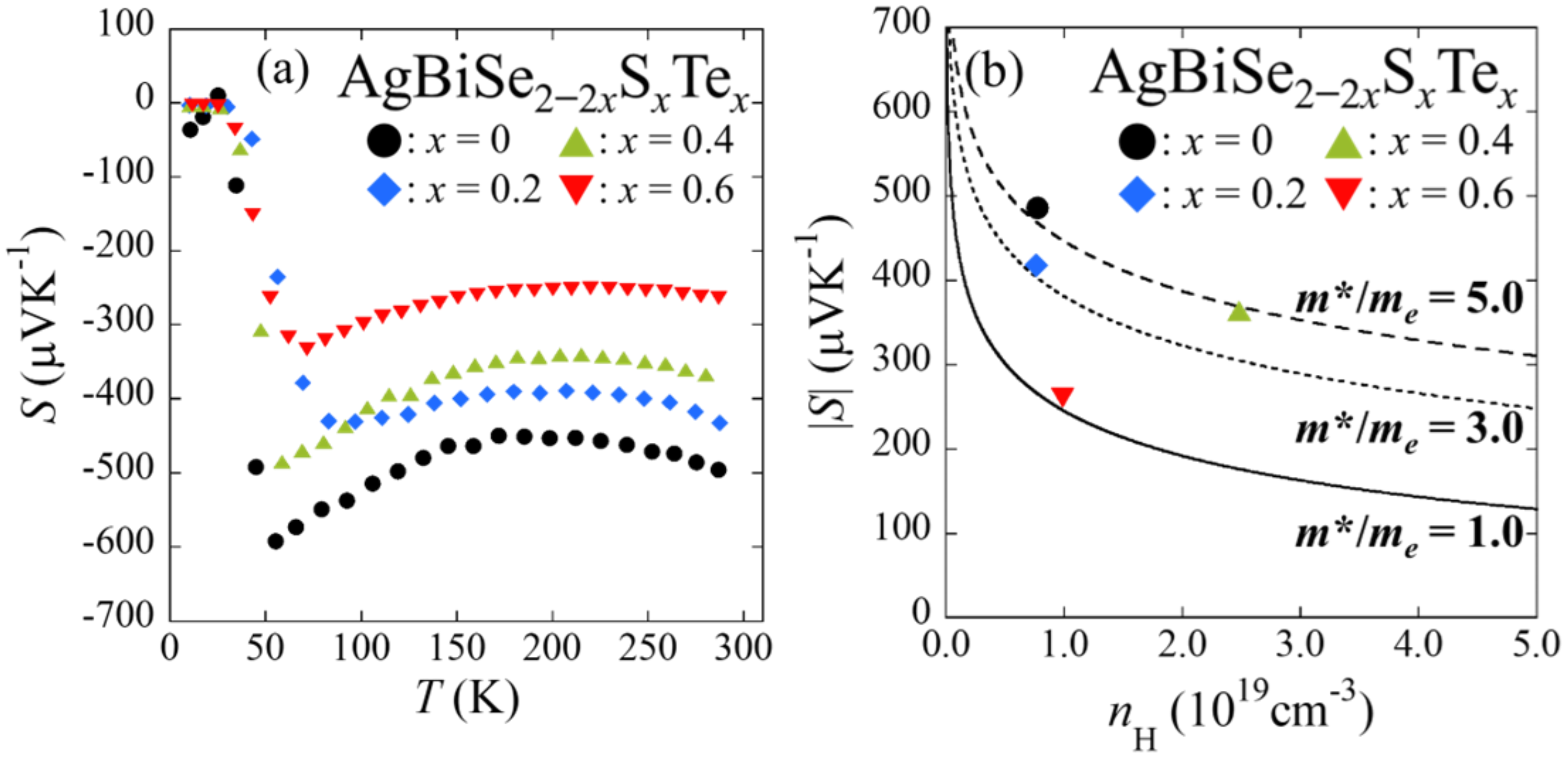

**Fig. 5** Temperature ($T$) dependence of (a) Seebeck coefficient ($S$). (b) Pisarenko plots for absolute $S$ ($|S|$) vs Hall carrier concentration ($n_H$) at 275 K.

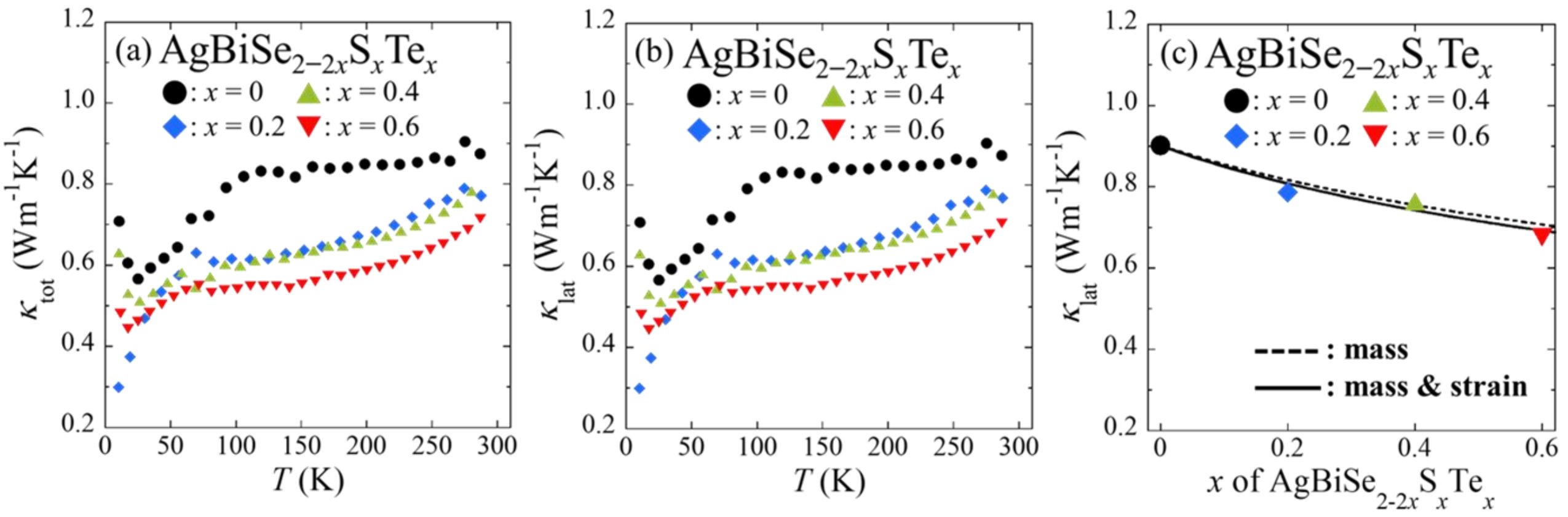

**Fig. 6** Temperature dependence of (a) total thermal conductivity ($\kappa_{tot}$) and (b) lattice thermal conductivity ($\kappa_{lat}$) for quenched $AgBiSe_{2-2x}S_xTe_x$ samples. (c) PDS model fitting to $\kappa_{lat}$ (color plots) at 275 K. The Dashed line shows mass contribution, and the solid line shows mass and strain contributions.

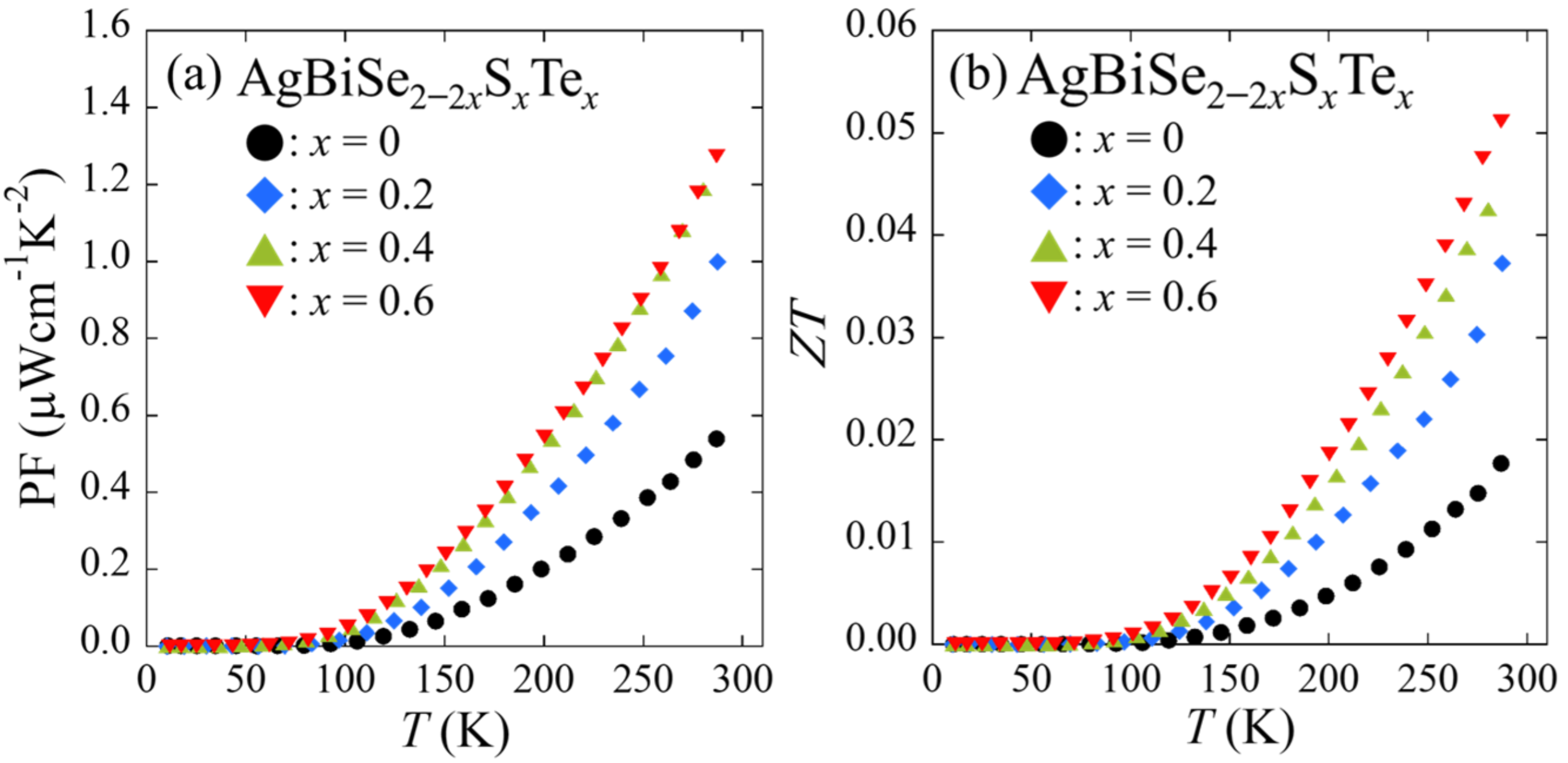

**Fig. 7** Temperature dependence of (a) PF and (b) *ZT* of quenched $AgBiSe_{2-2x}S_xTe_x$ samples.

**Table 1** Chemical composition, mixing entropy, lattice constant, volume, and activation energy of electrical resistivity for quenched $AgBiSe_{2-2x}S_xTe_x$ samples.

| $x$ | nominal | actual | $\Delta S_{mix}/R$ | $a$ (Å) | $V$ (Å$^3$) | Δ (meV) |
|---|---|---|---|---|---|---|
| 0.0 | $AgBiSe_2$ | $Ag_{1.02}Bi_{1.04}Se_{1.94}$ | 0.693 | 5.8269 | 197.84 | 62.2 |
| 0.2 | $AgBiSe_{1.6}S_{0.2}Te_{0.2}$ | $Ag_{1.04}Bi_{1.02}Se_{1.58}S_{0.15}Te_{0.21}$ | 1.403 | 5.8459 | 199.78 | 55.9 |
| 0.4 | $AgBiSe_{1.2}S_{0.4}Te_{0.4}$ | $Ag_{1.04}Bi_{1.03}Se_{1.17}S_{0.36}Te_{0.40}$ | 1.659 | 5.8692 | 202.18 | 51.3 |
| 0.6 | $AgBiSe_{0.8}S_{0.6}Te_{0.6}$ | $Ag_{1.02}Bi_{1.03}Se_{0.80}S_{0.57}Te_{0.57}$ | 1.764 | 5.8834 | 203.65 | 47.6 |

# Supporting Information

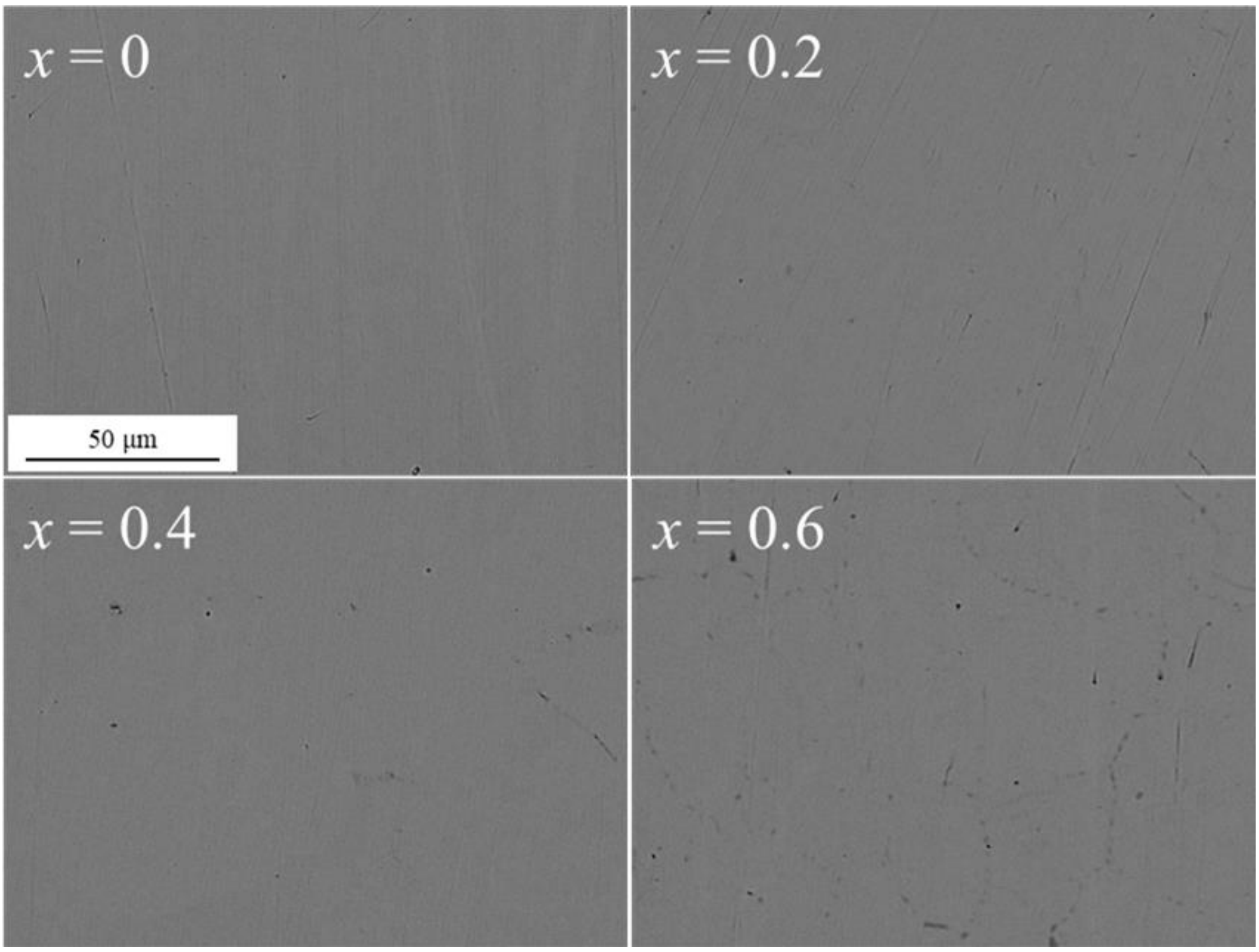

Fig. S1 SEM images for quenched $AgBiSe_{2-2x}S_xTe_x$ samples.

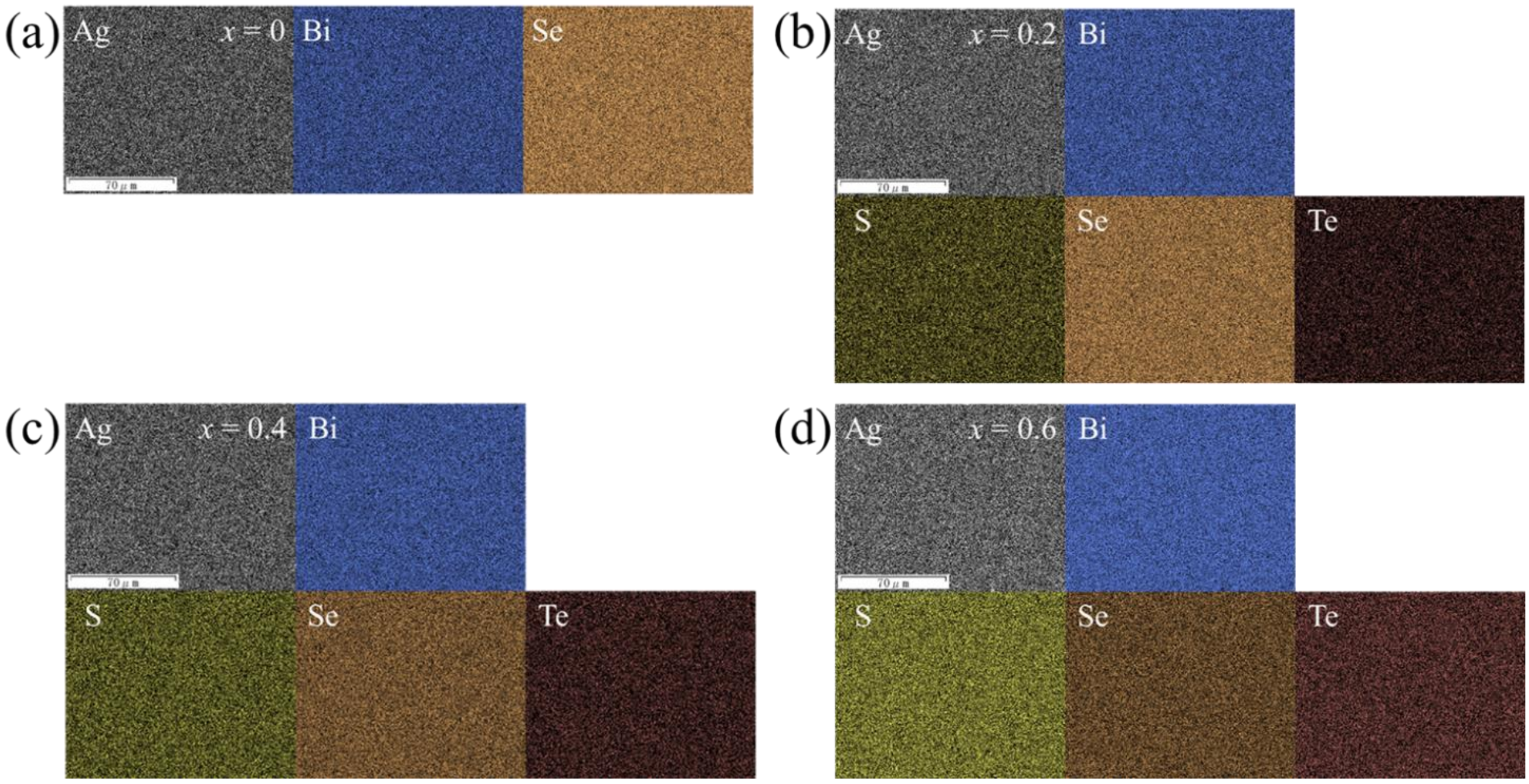

Fig. S2 EDX mapping for quenched $AgBiSe_{2-2x}S_xTe_x$ samples.

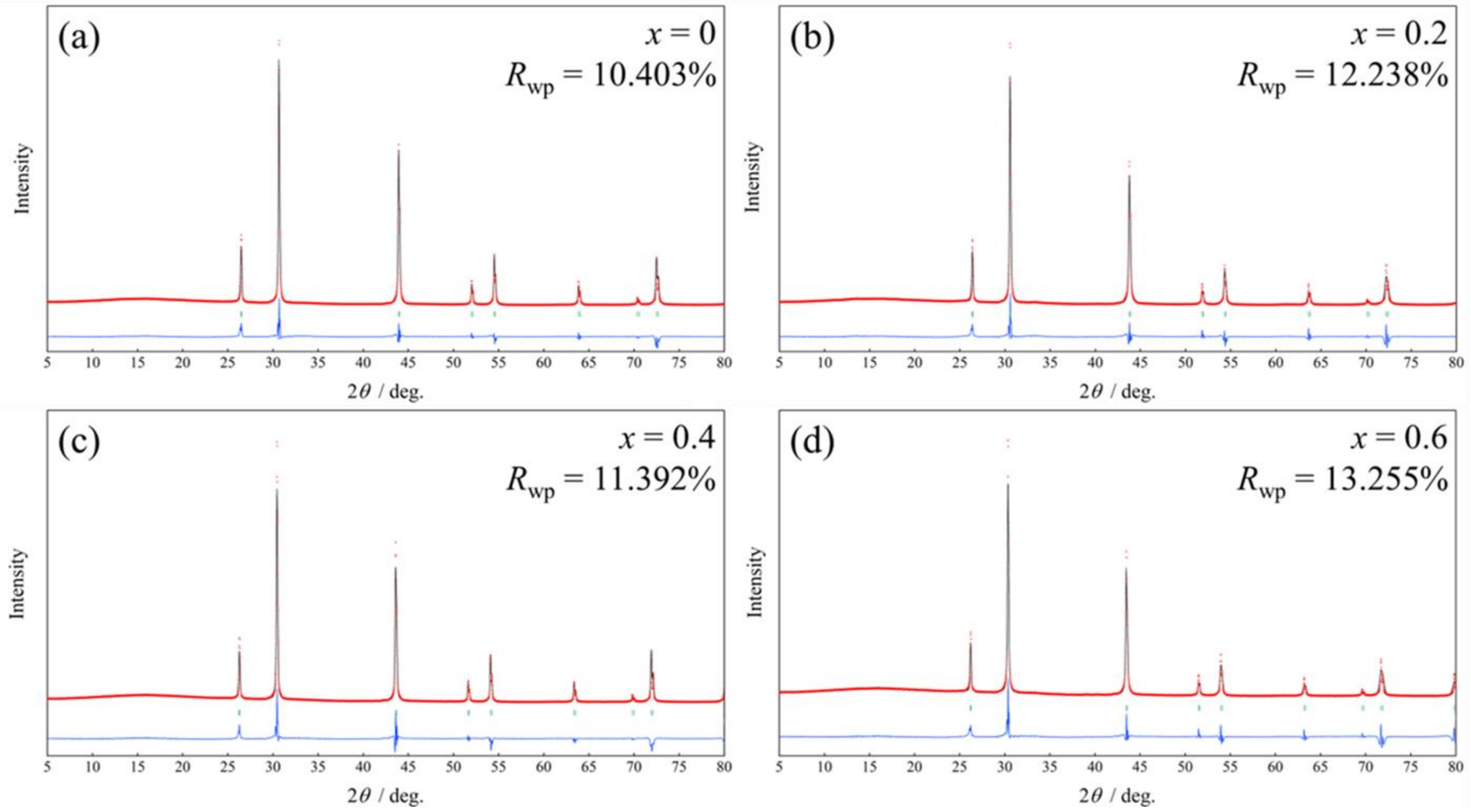

Fig. S3 The result of Rietveld refinement for quenched $AgBiSe_{2-2x}S_xTe_x$ samples.

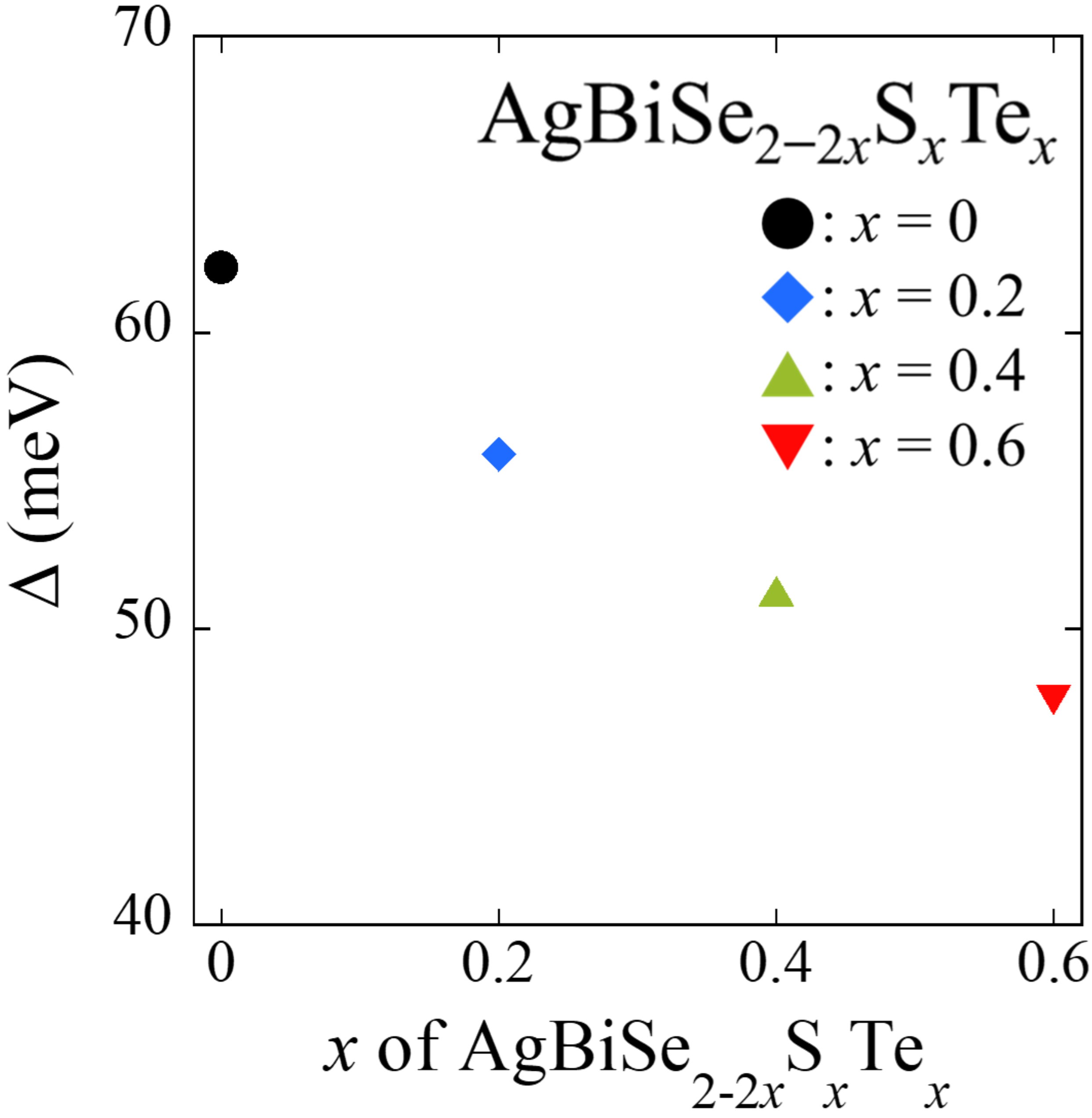

Fig. S4 Activation energy ( Δ ) of electrical resistivity for quenched $AgBiSe_{2-2x}S_xTe_x$ samples.

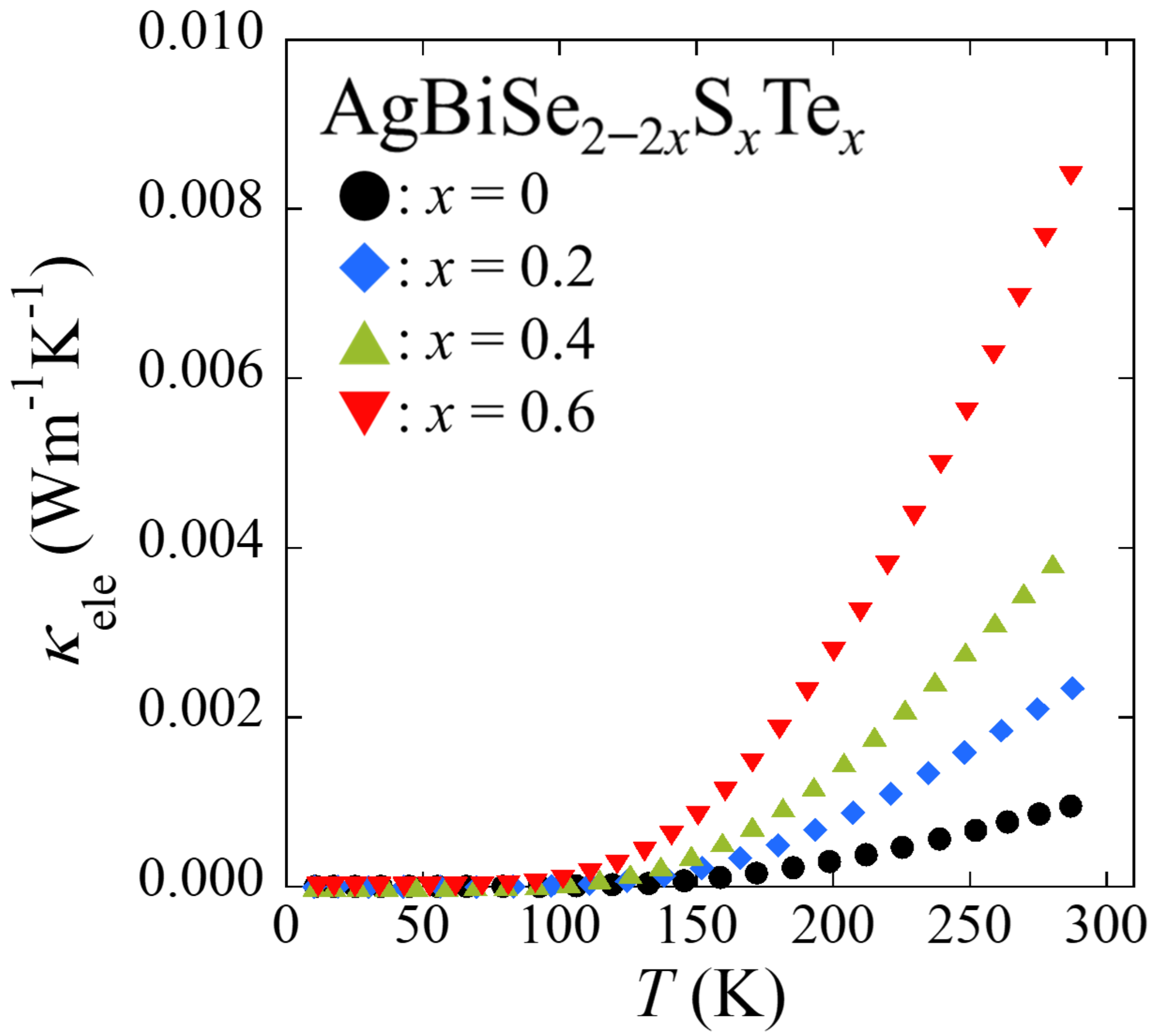

Fig. S5 Temperature ($T$) dependence of electronic thermal conductivity ($\kappa_{ele}$) for quenched $AgBiSe_{2-2x}S_xTe_x$ samples.

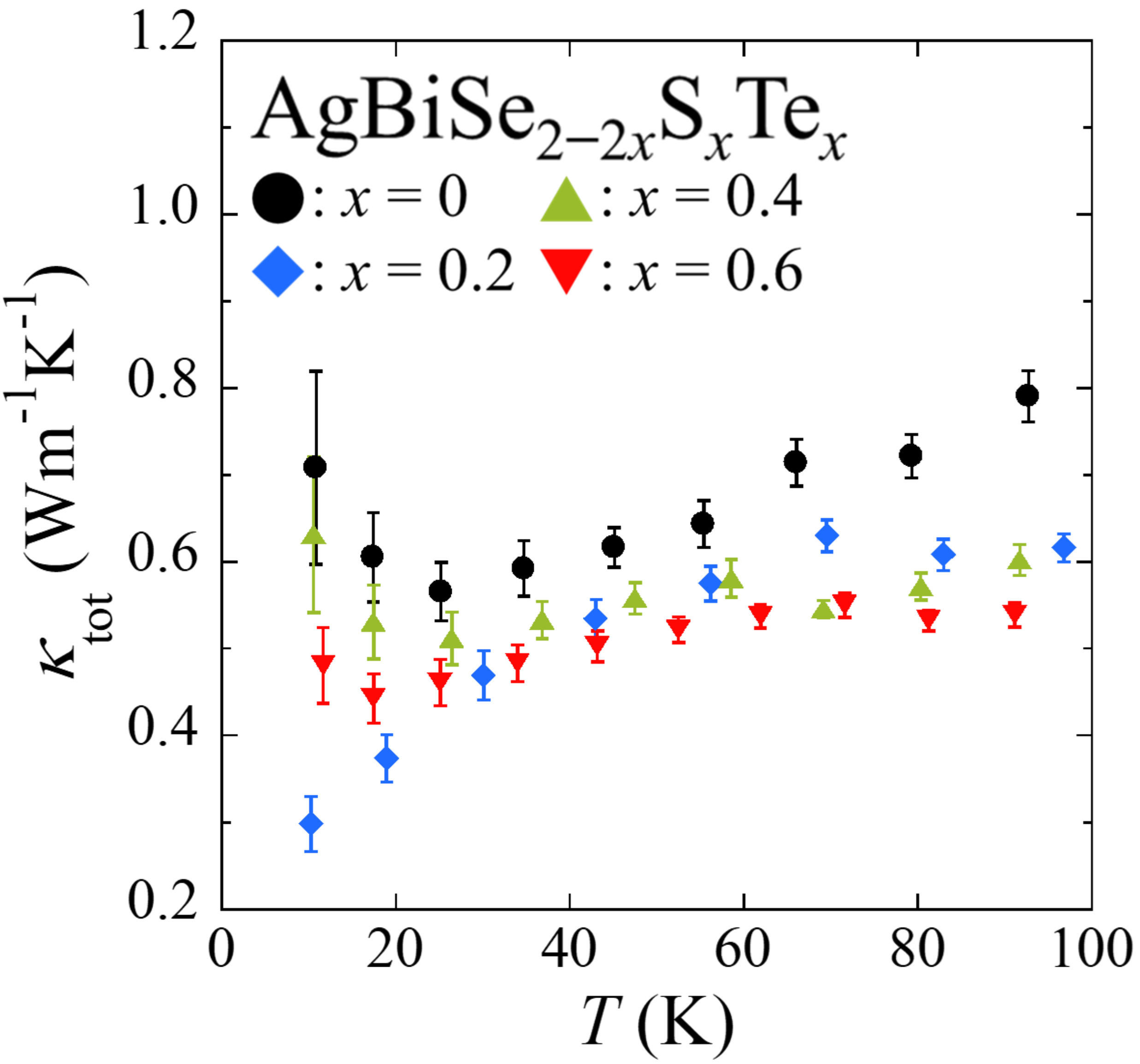

Fig. S6 Temperature dependence ($T$) of total thermal conductivity ($\kappa_{tot}$) with error bar for quenched $AgBiSe_{2-2x}S_xTe_x$ samples.